\RequirePackage[T1]{fontenc}
\documentclass[12pt,a4paper]{article}

\usepackage[height=8.85in,width=6.75in]{geometry}
\usepackage{amsmath}
\usepackage{mathrsfs}
\usepackage{amssymb}
\usepackage{mathtools}
\usepackage{slashed}
\usepackage{braket}
\usepackage{graphicx,rotating}
\usepackage{xcolor}
\usepackage[utf8]{inputenc}
\definecolor{bleudefrance}{rgb}{0.19, 0.55, 0.91}
\definecolor{desyblue}{HTML}{009EE2}
\definecolor{desyorange}{HTML}{FD8800}
\definecolor{dark_red}{rgb}{0.7, 0., 0.}
\definecolor{light_pink}{rgb}{1,0.4,0.4}
\definecolor{lblue}{rgb}{0.384602,0.117763,0.973947}
\usepackage[bookmarksopen,colorlinks=true,
linkcolor=lblue,
citecolor=lblue,
urlcolor=lblue,
linktoc=all]{hyperref}
\usepackage{cite}
\usepackage{tikz}
\usepackage{tikz-cd}
\usepackage{times}
\usepackage{courier}
\usepackage{bm,physics}
\usepackage{xcolor}
\usepackage{mdframed}
\usepackage{datetime}
\usepackage{dsfont}
\usepackage{multirow}
\usepackage{cancel}
\usepackage{caption}
\usepackage{subcaption}
\usepackage{graphicx}
\usepackage[compat=1.1.0]{tikz-feynman}
\usepackage{pifont}
\usepackage{romannum}
\AtBeginDocument{\pagenumbering{arabic}}
\usepackage{ulem}
\usepackage{float}

\newcommand{\rhoc}{\rho_{\mathrm{c}}}
\newcommand{\TRH}{T_{\mathrm{RH}}}
\newcommand{\Tdecay}{T_{\mathrm{decay}}}

\newcommand{\ee}{\mathrm{e}}
\newcommand{\GeV}{\mathrm{GeV}}
\newcommand{\diag}{\mathrm{diag}}
\newcommand{\Mpl}{M_\text{Pl}}

\newcommand{\calL}{\mathcal{L}}

\newcommand{\calO}{\mathcal{O}}

\newcommand{\calP}{\mathcal{P}}

\newcommand{\bae}[1]{\begin{align} #1 \end{align}}

\newcommand{\ainf}{a_\text{inf}}
\newcommand{\aRH}{a_\text{RH}}

\newcommand{\GW}{\mathrm{GW}}

\definecolor{monza}{HTML}{CF000F}
\definecolor{darkblue}{HTML}{00008b}
\definecolor{darkmagenta}{HTML}{8b008b}

\definecolor{pblue}{rgb}{0.252, 0.237, 0.882}
\definecolor{pgreen}{rgb}{0.271, 0.537, 0.200}
\definecolor{ppink}{rgb}{0.959, 0.091, 0.812}

\numberwithin{equation}{section}

\begin{document}
\begin{titlepage}

\begin{flushright}
\end{flushright}

\vskip 3cm

\begin{center}	
	{\Huge \bfseries Gravitational waves from graviton
 \\
 	\vskip .3cm
  Bremsstrahlung with kination phase
}	
	\vskip 1.5cm
  {\large
  Ryoto Inui$^{\ast,}$\footnote{\href{inui.ryoto.a3@s.mail.nagoya-u.ac.jp}{inui.ryoto.a3@s.mail.nagoya-u.ac.jp}}, 
  Yusuke Mikura$^{\ast,}$\footnote{\href{mikura.yusuke.s8@s.mail.nagoya-u.ac.jp}{mikura.yusuke.s8@s.mail.nagoya-u.ac.jp}}, 
  and Shuichiro Yokoyama$^{\dag, \ast, \ddag,}$\footnote{\href{shu@kmi.nagoya-u.ac.jp}{shu@kmi.nagoya-u.ac.jp}}
  }

  \vskip 1.5cm

 \def\arraystretch{1}
	\begin{tabular}{ll}
		$^\ast$& Department of Physics, Nagoya University, 
  \\& Furo-cho Chikusa-ku, Nagoya, Aichi 464-8602, Japan\\
		$^\dag$& Kobayashi-Maskawa Institute, Nagoya University, 
    \\& Furo-cho Chikusa-ku, Nagoya, Aichi 464-8602, Japan\\
		$^\ddag$& Kavli IPMU (WPI), UTIAS, The University of Tokyo,
    \\& 5-1-5 Kashiwanoha, Kashiwa, Chiba 277-8583, Japan\\
	\end{tabular}
	
    \vskip 1.5cm
	
\end{center}

\noindent
Gravitational waves (GWs) from gravitational three-body decay (graviton Bremsstrahlung process) can leave an indelible signal at ultrahigh frequencies. We focus on a scenario where superheavy particles are produced gravitationally at a transition between the inflationary and kination phases and analyze the detectability of the signal in the presence of GWs generated from the vacuum fluctuations during inflation.
We find that, in many cases, GWs from the graviton Bremsstrahlung are buried in the stochastic gravitational wave background originating from inflation. However, if the superheavy particles are as heavy as the Planck scale, the graviton Bremsstrahlung can produce a sizable amount of GWs, surpassing the inflationary ones.

\end{titlepage}

\setcounter{tocdepth}{3}
\tableofcontents

\section{Introduction}
The discovery of gravitational waves (GWs) by the LIGO-Virgo collaborations~\cite{LIGOScientific:2016aoc, LIGOScientific:2017vwq} has opened a new era where GWs can be used as a practical tool to understand the evolution of the universe as well as the underlying fundamental physics. 
Following the detection of GWs from mergers of compact binaries by the ground-based experiments, the international pulsar timing array (PTA) networks recently reported evidence of stochastic gravitational wave background~\cite{NANOGrav:2023gor, EPTA:2023fyk, Reardon:2023gzh, Xu:2023wog}. While it is considered that the signal stems from the superposition of GWs from supermassive black hole binaries, there exist several interpretations originating in cosmological phenomena~\cite{NANOGrav:2023hvm} and the report made it more realistic to utilize GWs to understand details of the universe.

GWs with cosmological origin appear over a wide range of frequencies. The most notable imprint on all scales comes from the primordial tensor perturbations generated quantum-mechanically during cosmic inflation~\cite{Starobinsky:1980te, Sato:1980yn, Guth:1980zm, Linde:1981mu, Albrecht:1982wi, Linde:1983gd}. 
The detection of such inflationary GWs allows us to probe particle contents and their interactions at presumably the highest energy scale accessible to us. 
In addition to the inflationary GWs, a sizable amount of cosmological GWs can be produced by several phenomena, e.g. the first-order phase transition~\cite{Apreda:2001us, Grojean:2006bp, Espinosa:2008kw, Caprini:2015zlo, Kakizaki:2015wua, Hashino:2016rvx, Hashino:2016xoj} and the formation of topological defects~\cite{Vilenkin:1981bx, Vachaspati:1984gt, Caldwell:1991jj, Damour:2000wa, Damour:2001bk, Hiramatsu:2013qaa}. The signals with such origins typically appear from a few mHz to a few hundreds of Hz corresponding to the horizon size at the formation epoch and have the potential to be tested by forthcoming observations such as LISA~\cite{LISA:2017pwj}, the Einstein Telescope (ET)~\cite{Punturo:2010zz, Hild:2010id, Sathyaprakash:2012jk, Maggiore:2019uih}, the Big Bang Observer (BBO)~\cite{Crowder:2005nr, Corbin:2005ny, Harry:2006fi}, and DECIGO~\cite{Seto:2001qf, Kudoh:2005as}. 

Recently, reheating has been attracting some attention as a possible source of cosmological GWs. Not only the scattering of the inflaton~\cite{Ema:2020ggo,Klose:2022knn} but also inevitable gravitational processes \cite{Nakayama:2018ptw, Huang:2019lgd, Barman:2023ymn, Barman:2023rpg, Kanemura:2023pnv, Bernal:2023wus, Hu:2024awd} may generate a noticeable amount of GWs. Interestingly, the spectrum of GWs generated during the reheating phase tends to have a peak at high frequencies, expecting that the inflationary scale is sufficiently high. As high-frequency GWs may offer an arena with good opportunities to investigate new physics without astrophysical interference, it is of great importance to scrutinize possible realizations of GWs during reheating.

Among gravitational interactions, three-body decay called graviton Bremsstrahlung has been a process in question. This process occurs whenever a parent particle decays into light particles through two-body decay. However, the amplitude of the produced GWs is suppressed when the parent particle is identified as the inflaton because the branching ratio of the three-body decay is suppressed by the ratio $M/\Mpl$ where $M$ is the mass of the parent particle and $\Mpl$ is the reduced Planck mass~\cite{Nakayama:2018ptw, Huang:2019lgd, Barman:2023ymn, Bernal:2023wus, Hu:2024awd}. Therefore, it is interesting to look for scenarios in which superheavy particles are naturally produced in the early universe as investigated in Ref.~\cite{Hu:2024awd}.

In this work, we discuss GW signatures from gravitationally produced superheavy particles. In the case of kinetically-driven models~\cite{Armendariz-Picon:1999hyi, Garriga:1999vw} or in some potential-driven models, the universe undergoes the kination phase~\cite{Kunimitsu:2012xx, Giovannini:1999bh, Giovannini:1999qj, Babusci:1999ky, Riazuelo:2000fc, Tashiro:2003qp, Sami:2004xk, Artymowski:2017pua, Gouttenoire:2021jhk} in which the kinetic energy of a scalar field such as inflaton dominates the total energy budget. In this circumstance, massive particles can be produced through gravitational particle production~\cite{Ford:1986sy, Chun:2009yu, Ford:2021syk, Kolb:2023ydq} and reheating is completed through the decay of the massive particles into radiation.
The massive particles produced in this way can be heavier than the scale of inflation depending on details of a transition between the inflation and kination phases~\cite{Hashiba:2018iff} so that the graviton Bremsstrahlung process can give rise to a significant amount of GWs. In the presence of the kination phase, one thing to be taken into account is that the inflationary GWs observed today are relatively enhanced at high frequencies~\cite{Giovannini:1999bh, Giovannini:1999qj, Babusci:1999ky, Riazuelo:2000fc, Tashiro:2003qp, Sami:2004xk, Artymowski:2017pua, Mikura:2021ldx}. This is due to the quick decay of the background energy density $\rho_{\mathrm{inf}}\propto a^{-6}$, while radiation decays a bit slower as $\rho_{\mathrm{R}}\propto a^{-4}$. This enhancement makes it less obvious whether or not we can extract pure information about new physics during reheating with the use of high-frequency GWs, which is the topic covered in the paper.

The paper is organized as follows. In Sec.~\ref{sec. Setup and the background dynamics}, we first define our model and present two possible dynamics of our universe. Then, in Sec~\ref{Sec. Gravitational Waves}, we discuss the detectability of GWs produced by the graviton Bremsstrahlung in the presence of the inflationary GWs enhanced by the kination phase. Finally, we conclude in Sec.~\ref{sec. discussions}.
Throughout the paper, we adopt the natural unit $c=\hbar=1$ and $\eta_{\mu\nu} = \diag~(-1,1,1,1)$ is used for the sign of the Minkowski metric.

\section{Setup and the background dynamics}\label{sec. Setup and the background dynamics}
As mentioned in the introduction, massive particles can be gravitationally produced due to the consequence of a change of two vacua between the inflation to kination phases with the breaking of conformal symmetry. This is clear if one considers a massive particle $S$ which couples non-minimally to gravity:
\bae{
S = \int \dd^4x \sqrt{-g}\left[\frac{\Mpl^2}{2}R + \frac{1}{2}\xi S^2 R -\frac{1}{2}(\partial S)^2 -\frac{1}{2}M^2 S^2 \right] ~.
}
It is easy to derive that a (normalized) mode function of the scalar $\chi_k$ satisfies 
\bae{\label{Eq mode function}
\frac{\dd^2 \chi_k}{\dd \eta^2} +\left[k^2+a^2M^2+\left(\xi+\frac{1}{6}\right)\frac{a^{\prime\prime}}{a}\right]\chi_k=0 ~,
}
where $k$ is the comoving wavenumber, $a$ is the scale factor and primes denote derivatives with respect to conformal time $\eta$. In the limit of $M\to 0$ and $\xi \to -1/6$, the mode equation takes a simple form with a time-independent frequency, meaning that particle production does not occur. 
While the particle production can be sourced by both the mass term and non-minimal coupling, the symmetry breaking by the mass term has been well studied~\cite{Mamaev:1976zb, Birrell:1979pi, Turner:1987vd, Chung:2001cb, Kannike:2016jfs}, which may be because the conformal coupling with $\xi=-1/6$ appears in some UV models based on supergravity and superstring. Recently, this direction was developed analytically and numerically in Ref.~\cite{Hashiba:2018iff} with the aim of studying the production of superheavy particles.

In this work, we consider a similar setup studied in Ref.~\cite{Hashiba:2018iff} for analytical purposes. Concretely, we consider the action of the conformally coupled superheavy particle $S$ with a trilinear coupling between the superheavy particle $S$ and a complex doublet $\Phi$, where the Lagrangian is defined by~\footnote{A key ingredient for gravitational particle production is the breaking of conformal symmetry. Thus one can consider gravitational production for massive Dirac fermions as the mass term breaks the symmetry~\cite{Hashiba:2022bzi}.}
\bae{
\calL = \calL_{\mathrm{inf}} -\frac{1}{12}S^2 R -\frac{1}{2}(\partial S)^2-\frac{1}{2}M^2 S^2 - \partial \Phi^\dag \partial \Phi - m^2 \Phi^\dag \Phi - \mu S \Phi^\dag \Phi ~,
}
where $\mu$ is a dimensionful coupling, and $M$ and $m$ are masses of $S$ and $\Phi$, respectively. Here we do not specify a concrete form of the inflationary sector $\calL_{\mathrm{inf}}$. The post-inflationary dynamics is determined by the amount of the energy density of the produced superheavy particles and the time of decay into light fields which create the thermal bath for the successful Big Bang nucleosynthesis (BBN). The energy density of the superheavy particles at the production is given by~\cite{Hashiba:2018iff}
\bae{\label{Eq. energy superheavy at production}
\rho_{S, \mathrm{inf}} \simeq 2\times 10^{-4} ~ \ee^{-4M\Delta t} M^2 H_{\mathrm{inf}}^2 ~,
}
where $H_\mathrm{inf}$ is the Hubble scale at the end of inflation, or in other words the moment of gravitational particle production, with a subscript ``inf'' and $\Delta t$ is the time scale of transition which is typically the Hubble time but can be small in some occasions.
The time of decay is controlled by the decay rate $\Gamma^{(2)}$ of a process $S \to \Phi^\dag \Phi$, given by
\bae{\label{eq-two-body-decay}
\Gamma^{(2)} = \frac{\mu^2}{8\pi M}\sqrt{1-4y^2} ~, \quad   y \coloneqq \frac{m}{M} ~,
}
where we assume throughout that the field $\Phi$ is enough light $y \ll 1$ so that it can be treated as radiation.~\footnote{In the calculation of GWs, we take $y = 10^{-5}$.}

\begin{figure}
    \centering
            \centering
            \includegraphics[width=0.55\hsize]{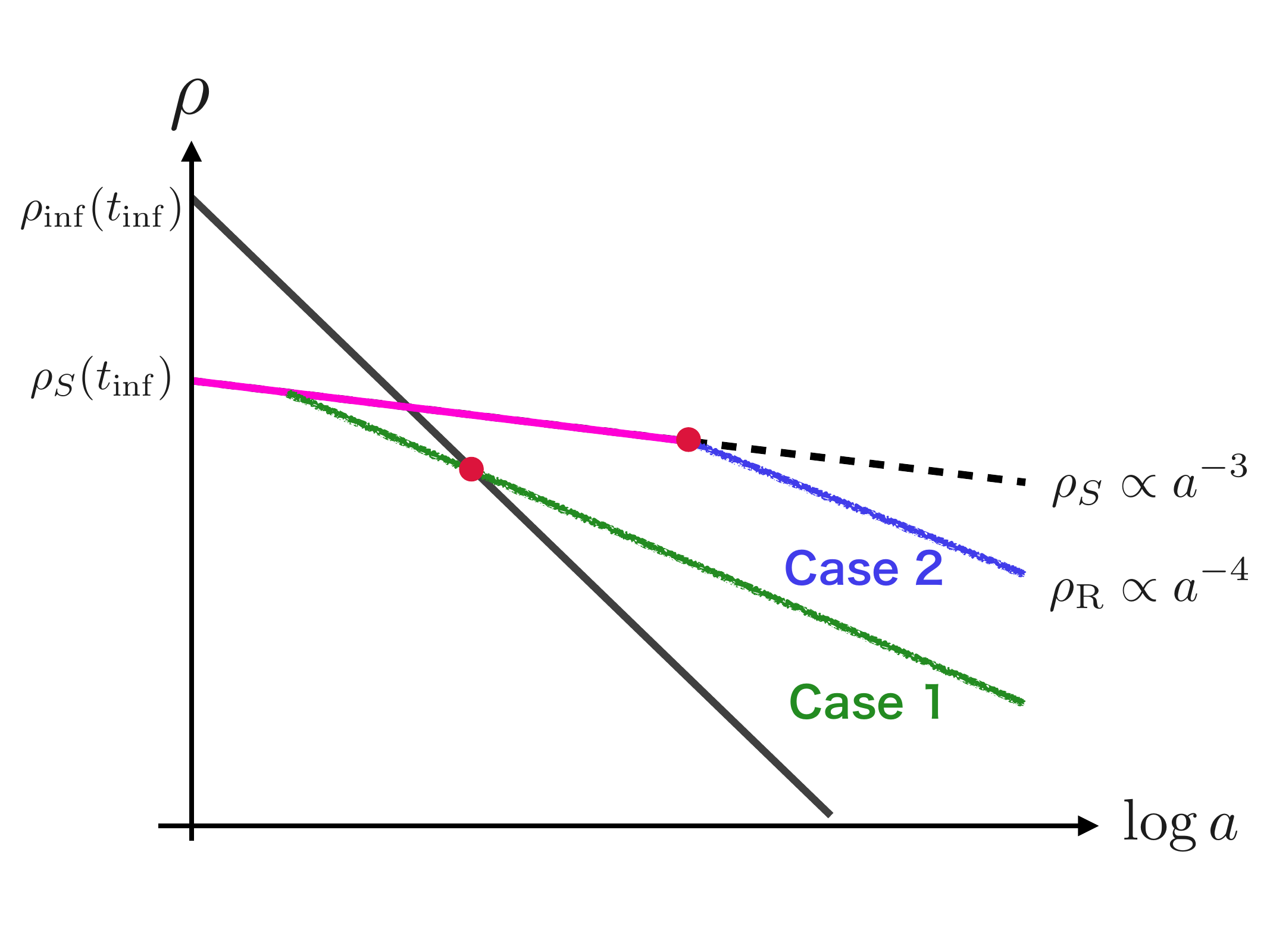}
    \caption{The evolution of energy densities. The black solid line represents the inflaton's energy density which decays as $\rho_{\mathrm{inf}}\propto a^{-6}$. The Pink line describes the evolution of the superheavy particles which evolve as $\rho_{\mathrm{S}}\propto a^{-3}$. Depending on when the superheavy particles decay into radiation, there exist two different cases. In case~1, the decay occurs before the superheavy particles dominate the universe. In case~2, the universe experiences an intermediate matter-dominated universe between the kination and radiation-dominated phases. Red blobs denote the end of reheating in each case.} 
    \label{Fig. Image}
\end{figure}

For simplicity, here we assume that the inflaton's energy density evolves as $\rho_{\mathrm{inf}} \propto a^{-6}$ during 
the kination phase without entropy injection and the superheavy particles decay into radiation instantaneously. Then, due to the difference in the time evolution of energy densities of the superheavy particles and the radiation, there exist two background evolution paths that the universe follows,
depending on when the superheavy particles decay into radiation. 
In the first case, the superheavy particles decay into radiation before the energy density of the superheavy particles exceeds the one of the inflaton. It follows that the end of reheating is defined at the moment when $\rho_{\mathrm{R}}=\rho_{\mathrm{inf}}$ is satisfied.
We call this path case~1.
On the other hand, in the second case, the universe undergoes an intermediate matter-dominated phase between the kination and radiation-dominated phases. The end of reheating is evaluated when the superheavy particles decay into radiation. We call this path case~2. The dynamics of the energy densities are schematically described in Fig.~\ref{Fig. Image}.

\subsection{Critical case}
Let us first find a set of critical parameters which define the two cases. These parameters can be obtained by assuming the superheavy particles decay when their energy density satisfies $\rho_{S}= \rho_{\mathrm{inf}}$.
Recalling that the superheavy particles are non-relativistic and taking the evolution of the scale factor into account, the reheating temperature in this critical case $T_{\ast}$ can be estimated by using Eq.~\eqref{Eq. energy superheavy at production} as~\cite{Hashiba:2018iff} 
\bae{\label{eq-critical-temperature}
T_{\ast} \simeq 2 \times 10^{13} ~ \ee^{-2M\Delta t}\left(\frac{M}{\Mpl}\right)\left(\frac{H_{\mathrm{inf}}}{10^{13} ~\GeV}\right)^{1/2} \GeV ~,
}
where we use $g_\ast=106.75$ for the number of relativistic degrees of freedom at high temperature and again $\Mpl =2.4\times 10^{18} ~\mathrm{GeV}$ denotes the reduced Planck scale. As can be seen in the Fig.~\ref{Fig. Image}, this critical case provides the maximum reheating temperature. Assuming that reheating is instantaneously completed, one can derive a critical value of the coupling $\mu$ using a relation
\bae{\label{eq-critical-decay}
T_{\ast} \simeq \left(\frac{90}{2\pi^2 g_\ast}\right)^{1/4}\sqrt{\Gamma^{(2)}\Mpl} ~.
}
Using Eqs.~\eqref{eq-two-body-decay}, \eqref{eq-critical-temperature}, and \eqref{eq-critical-decay}, we find that a critical value of the coupling $\mu$ in the limit of $y \to 0$ is approximately given by
\bae{
\mu_{\ast } \simeq 2 \times 10^{14} ~ \ee^{-2M\Delta t}\left(\frac{M}{\Mpl}\right)^{3/2} \left(\frac{H_{\mathrm{inf}}}{10^{13} ~\GeV}\right)^{1/2} \GeV ~.
\label{eq-critical-coupling}
}
\subsection{Case~1: decay before the domination $(\mu > \mu_*)$}
Let us next consider the case~1 where the superheavy particles decay into radiation before they dominate the universe. Letting $\aRH$ be the scale factor at the end of reheating, the reheating temperature $\TRH$ can be obtained by solving 
\bae{
\frac{\pi^2 g_\ast }{30}\TRH^4 = 3\Mpl^2 H_{\mathrm{inf}}^2 \left(\frac{\aRH}{\ainf}\right)^{-6} ~.
}
The scale factor at the end of reheating can be written in terms of the energy densities as
\bae{
\frac{\aRH}{a_{\mathrm{inf}}} = \left(\frac{\rho_{\mathrm{R, \mathrm{decay}}}}{\rho_{S, \mathrm{inf}}}\right)^{-1/3}\left(\frac{\rho_{\mathrm{R, \mathrm{RH}}}}{\rho_\mathrm{R, \mathrm{decay}}}\right)^{-1/4} ~,
}
where the subscript ``decay'' denotes the moment of decay of the superheavy particles into radiation. With this relation, the reheating temperature can be expressed as a function of the parameters of the theory as
\bae{\label{eq:caseI_reheating}
\TRH & \simeq \frac{1}{\sqrt{3}}\left(\frac{\pi^2 g_\ast}{30}\right)^{-1/4}\frac{\rho_{S, \mathrm{inf}}^{3/4}}{\Mpl H_{\mathrm{inf}}^{3/4}}\left(\Gamma^{(2)}\right)^{-1/4} ~, \nonumber
\\
& \simeq  1 \times 10^{13} ~ \left(\frac{10^{15} ~\GeV}{\mu}\right)^{1/2}\ee^{-3M\Delta t}\left(\frac{M}{\Mpl}\right)^{7/4} \left(\frac{H_{\mathrm{inf}}}{10^{13}~\GeV}\right)^{3/4} ~ \GeV ~,
}
where we again assume that the number of relativistic degrees is unchanged during reheating. Note that the coupling $\mu$ has to be stronger than the critical value defined by Eq.~\eqref{eq-critical-coupling} with given $M$ and $H_{\mathrm{inf}}$.
Note also that the coupling is bounded from above to realize the BBN, which requires 
\bae{
\mu < 10^{47}\,\ee^{-6M\Delta t}\left(\frac{M}{\Mpl}\right)^{7/2} \left(\frac{H_{\mathrm{inf}}}{10^{13}~\GeV}\right)^{3/2} ~ \GeV ~.
}
If we are interested in superheavy particles and the energy scale of inflation is not too low, this bound is less restrictive compared to an EFT argument $\mu < \Mpl$.
\subsection{Case~2: decay after the domination $(\mu < \mu_*)$}
In the case~2, the universe undergoes the matter-dominated phase after kination. Then decay of the superheavy particles into radiation determines the end of reheating, where the reheating temperature is a solution of
\bae{
\frac{\pi^2 g_\ast}{30} \TRH^4  = 3 \Mpl^2 \left(\Gamma^{(2)}\right)^2 ~,
}
leading to
\bae{
\TRH  \simeq  \left(\frac{90}{\pi^2 g_\ast }\right)^{1/4}\sqrt{\Gamma^{(2)}\Mpl}
~\simeq 1 \times 10^{12} \left( \frac{\mu}{10^{13}~\GeV} \right) 
\left(\frac{M}{\Mpl}\right)^{-1/2}~\GeV
~.
}
Note that there exist the maximum and minimum values the coupling $\mu$ can take. The maximum value is trivially fixed by the critical coupling while the minimum value is determined by requiring that the BBN successfully occurs. The second condition roughly means $\TRH > 1 ~ \mathrm{MeV}$, so that the coupling is constrained to be 
\bae{
10^{-2} \left(\frac{\Mpl}{M}\right)^{1/2} ~\GeV <\mu <\mu_\ast ~.
}
\subsection{Comment on gravitationally produced gravitons}
Through the gravitational particle production, massless spin-2 gravitons are also produced and they behave as freely propagating radiation. Their energy density at the production is twice as much as that of a minimally coupled massless scalar field, which is given by~\cite{Kunimitsu:2012xx, Hashiba:2018tbu}
\bae{
\rho_{\mathrm{GW, \mathrm{inf}}}^{\mathrm{grav}}\simeq c ~ \frac{9}{16\pi^2} H_{\mathrm{inf}}^4 ~,
}
where $c$ is a numerical factor of order $\calO (1)$ or $\calO (10)$ depending on details of the transition. At the time of production, the energy density of the gravitationally produced gravitons is smaller than that of the superheavy particles if the mass $M$ satisfies
\bae{\label{eq-condition-GWs-sub}
M \gtrsim 20 ~ c^{1/2} ~ \ee^{2M\Delta t} H_{\mathrm{inf}} ~.
}
Assuming that the mass of the superheavy particles is almost the same as the inverse of the transition time scale, $M\sim (\Delta t)^{-1}$, the above condition reads
\bae{
M \gtrsim 10^2 H_{\mathrm{inf}} ~.
}
Note that, if the condition~\eqref{eq-condition-GWs-sub} is not satisfied, the evolution of the universe can be complicated as a major component of the total energy density differs case by case. Since we are interested in GWs from the superheavy particles in this work, for simplicity we consider the parameter region where the condition~\eqref{eq-condition-GWs-sub} is satisfied.
\section{Gravitational wave spectrum}\label{Sec. Gravitational Waves}
We have seen that the universe follows different evolution paths depending on the parameters of the theory. The aim of this section is to compare observational imprints of the inevitably-produced GWs through the graviton Bremsstrahlung process and inflationary GWs.
In Sec.~\ref{Sec. Gravitational Bremsstrahlung}, we review the basics of GWs from the graviton Bremsstrahlung. After a few comments on other possible sources of GWs in Sec.~\ref{Sec. Other relevant sources}, we turn to examples of observational signatures in Sec.~\ref{Sec. Observational signatures}.

\subsection{Graviton Bremsstrahlung}\label{Sec. Gravitational Bremsstrahlung}
In any particle decay processes, a collection of the gravitons is inevitably emitted through the interaction
\bae{\label{Eq-action-int}
S_{\mathrm{int}} = \frac{1}{\Mpl}\int \dd^4x \sqrt{-g} ~ h^{\mu\nu} T_{\mu\nu} ~,
}
where $T_{\mu\nu}$ denotes the energy-momentum tensor and $h_{\mu\nu}$ is the graviton expanded around the Minkowski spacetime, defined by
\bae{
T_{\mu\nu} = -\frac{2}{\sqrt{-g}}\frac{\delta S_{\mathrm{matter}}}{\delta g^{\mu\nu}} ~, 
\quad 
g_{\mu\nu} \simeq \eta_{\mu\nu}+\frac{2}{\Mpl}h_{\mu\nu} ~.
}
The interaction~\eqref{Eq-action-int} gives rise to three-body decay $S\to \Phi^\dag\Phi+h_{\mu\nu}$ called the graviton Bremsstrahlung process and the total decay width can be written as 
\bae{
\Gamma \simeq \Gamma^{(2)} + \Gamma^{(3)} ~,
}
where $\Gamma^{(2)}$ and $\Gamma^{(3)}$ denote the widths of two-body and three-body decays, respectively. 
The differential decay width of the graviton Bremsstrahlung process can be calculated by integrating the associated amplitude for the three-body phase space without performing the integration for energy. Its explicit form is given by~\cite{Barman:2023rpg}
\bae{
\frac{\dd \Gamma^{(3)}}{\dd \ln E} = \frac{1}{32\pi^3}\frac{\mu^2}{M}\frac{M^2}{\Mpl^2}\left[\frac{(1-2x)(1-2x+2y^2)}{4 \alpha^{-1}}+ y^2 (y^2+2x-1)\ln \left(\frac{1+\alpha}{1-\alpha}\right)\right] ~,
}
with 
\bae{
x\coloneqq \frac{E}{M} ~, \quad \alpha \coloneqq \sqrt{1-\frac{4y^2}{1-2x}} ~,
}
where $E \lesssim M/2$ is the energy of the graviton.
 
The current density parameter of GWs is defined by 
\bae{
\Omega_{\GW} (f) \coloneqq \frac{1}{\rho_{{\rm c},0}}\eval{\frac{\dd \rho_\GW}{\dd \ln f}}_{T=T_0} =\Omega^{(0)}_\gamma \frac{g_{\ast, \mathrm{decay}}}{g_{\ast, 0}}\left(\frac{g_{\ast s, 0}}{g_{\ast s, \mathrm{decay}}}\right)^{4/3} \frac{\dd}{\dd \ln E}\left(\frac{\rho_{\GW, \mathrm{decay}}}{\rho_{\mathrm{R}, \mathrm{decay}}}\right) ~,
}
where a subscript “0” denotes the current value, $f$ denotes the frequency evaluated today, $g_{\ast, s}$ is the effective degrees of freedom for entropy, and $\rhoc$ is the critical energy density of the universe. As in the case of the radiation production through the decay of the superheavy particles, we assume that the graviton emission occurs instantaneously at the decay.
By utilizing the Boltzmann equations which describe the evolution of the energy densities
\bae{
\frac{\dd \rho_{S}}{\dd t} + 3 H \rho_{S} & = -\left(\Gamma^{(2)} +\Gamma^{(3)} \right)\rho_S ~,
\\
\frac{\dd \rho_{\mathrm{R}}}{\dd t} + 4 H \rho_{\mathrm{R}} & = \Gamma^{(2)}\rho_S +\int \frac{\dd \Gamma^{(3)}}{\dd \ln E} \left(1-\frac{E}{M}\right)\rho_S ~ \dd \ln E ~,
\label{eq-Friedman-rad}
\\
\frac{\dd \rho_{\GW}}{\dd t} + 4 H \rho_{\GW} & = + \int \frac{\dd \Gamma^{(3)}}{\dd \ln E}\frac{E}{M}\rho_S ~ \dd \ln E ~,
}
where
\bae{
\Gamma^{(3)} = \int \frac{ \dd \Gamma^{(3)}}{\dd \ln E} \dd \ln E ~,
}
one can relate the energy densities to the differential decay width as follows
\bae{\label{eq-rhoGW-Rad-int}
\frac{\dd}{\dd a}\left(\frac{\rho_{\GW}}{\rho_{\mathrm{R}}}\right) =
\frac{1}{aH}\frac{\rho_{S}}{\rho_{\mathrm{R}}}\left[
\int \frac{\dd \Gamma^{(3)}}{\dd E}\frac{E}{M}\dd E
-\frac{\rho_{\GW}}{\rho_{\mathrm{R}}}\int \frac{\dd \Gamma^{(3)}}{\dd E} \left(1-\frac{E}{M}\right)\dd E
-\frac{\rho_{\GW}}{\rho_{\mathrm{R}}}\Gamma^{(2)}
\right] ~.
}
Note that the second term in the bracket on the right-hand side would be negligible compared to the third term since a relation $\Gamma^{(3)} < \Gamma^{(2)}$ holds for $M<\Mpl$.
Then, in the limit of the instantaneous decay, Eq.~\eqref{eq-rhoGW-Rad-int} can be approximately integrated as
\bae{
  \frac{\dd}{\dd \ln E}\left(\frac{\rho_{\GW, \mathrm{decay}}}{\rho_{\mathrm{R}, \mathrm{decay}}}\right) \simeq \frac{1}{\Gamma^{(2)}}  \eval{\frac{\dd \Gamma^{(3)}}{\dd \ln E}\frac{E}{M}}_{\mathrm{decay}}
  ~.
}
We then arrive at an expression of the spectrum
\bae{\label{Eq-OmegaGWx}
\Omega_{\GW} = 5 \times 10^{-7}\left(\frac{M}{\Mpl}\right)^2 F (x,y) ~,
}
with
\bae{
F (x,y) \coloneqq \frac{x}{\sqrt{1-4 y^2}}\left[\frac{(1-2x)(1-2x+2y^2)}{4 \alpha^{-1}}+y^2 (y^2+2x-1)\ln \left(\frac{1+\alpha}{1-\alpha}\right)\right] ~.
}
Here we use $\Omega^{(0)}_\gamma \simeq 5.4 \times 10^{-5}$. It should be noted here that the right-hand side of Eq.~\eqref{Eq-OmegaGWx} is a function of the graviton energy at the production $E (=M x)$. This has to be rescaled to depict the spectrum $\Omega_{\GW} (f)$ in terms of observed frequency, which can be done by
\bae{
E = 2 \pi \frac{\Tdecay}{T_0} \left(\frac{g_{\ast s, \mathrm{decay}}}{g_{\ast s, 0}}\right)^{1/3} \times f ~.
}
Recalling that the temperature at decay is proportional to $\mu/\sqrt{M}$, the current density parameter of GWs from the graviton Bremsstrahlung for $E \lesssim M/2$ is proportional to
\bae{\label{Eq. mass-coupling dependence}
\Omega_{\GW} \propto M^2 x = \sqrt{M}\mu ~,
}
indicating that, from an observational point of view, it is preferable for the parent particles to have a heavy mass or strong coupling to matter fields.

\subsection{Other relevant sources}\label{Sec. Other relevant sources}
On top of GWs from the superheavy particle decay, there exist primordial GWs from inflationary tensor perturbations which have a non-trivial scaling in the presence of the kination phase. Letting $w$ be the equation-of-state parameter of the universe, the amplitude of the primordial GWs is roughly expressed as~\cite{Gouttenoire:2021jhk}
\bae{
\Omega_\mathrm{GW} \propto f^{-2\frac{1-3w}{1+3w}} \mathcal{P}_{T, {\rm inf}} ~,
}
showing that the amplitude grows linearly at high frequencies with $w=1$ corresponding to the kination phase. It is worth noting that the UV tail of the spectrum decays exponentially for frequencies larger than one corresponding to the mode that enters the horizon at the end of inflation~(recently pointed out in \cite{Pi:2024kpw}). The initial tensor power spectrum for inflationary GWs, denoted as $\mathcal{P}_{T, {\rm inf}}$, is simply given by 
\bae{
\mathcal{P}_{T,{\rm inf}} = \frac{2}{\pi^2}\left(\frac{H_{\mathrm{inf}}}{\Mpl}\right)^2\left(\frac{k}{k_\mathrm{p}}\right)^{n_t} ~,
}
where $k_\mathrm{p}$ is the pivot scale and $n_t$ is the spectral index. In slow-roll inflation, the spectral index can be written in terms of the tensor-to-scalar ratio $r$ as
\bae{
n_t\simeq -\frac{r}{8} ~.
}

It is widely known that GWs are also produced in the thermal plasma simply because particle scattering is accompanied by accelerations~\cite{Ghiglieri:2015nfa, Ghiglieri:2020mhm, Ringwald:2020ist}. Such GWs form a stochastic gravitational wave background which peaks at around $\calO(10)$ GHz range. We confirm that, in the current setup, GWs from the thermal plasma are smaller than GWs from the graviton Bremsstrahlung or inflation. Thus we neglect the contributions from the thermal plasma in this work.

\subsection{Observational signatures}\label{Sec. Observational signatures}
Let us now discuss the detectability of the signal from the graviton Bremsstrahlung process by comparing it with the inflationary GWs in the presence of the kination phase as the post-inflationary dynamics. For concreteness, without specifying an underlying inflationary model, we fix the tensor-to-scalar ratio to be $r\sim 10^{-2}$ which can be tested by the next-generation CMB experiments such as LiteBIRD~\cite{Hazumi:2019lys} and CMB-S4~\cite{CMB-S4:2016ple, CMB-S4:2020lpa}. The Hubble scale at the end of inflation is also fixed by using
\bae{
H_{\mathrm{inf}}^2\simeq \frac{\pi^2}{2}r\calP_\zeta\Mpl^2 ~,
}
where $\calP_\zeta \simeq 2.1\times 10^{-9}$ is the dimensionless power spectrum of curvature perturbation. For the other parameters $M$, $\Delta t$, and $\mu$, we consider four concrete cases listed in Table~\ref{Table. four cases}. Here we assume that
the time scale of transition is simply given by the inverse of the mass of the produced superheavy particles as an optimistic case where the particle production efficiently occurs. In addition, as an extremal case, we consider the case with $M=\Mpl$. As we have discussed in Sec.~\ref{sec. Setup and the background dynamics}, for fixed $M$ the background dynamics can be divided into two types depending on the choice of $\mu$.
\begin{table}[htbp]
\begin{center}
        \renewcommand{\arraystretch}{1.5}
    \begin{tabular}{|c|c|c|c|c|}
    \hline
   & $M = (\Delta t)^{-1}$ & $\mu_\ast$  & $\mu$ &$\TRH$
    \\
    \hline
     A-1 & $10^3 H_{\mathrm{inf}}$ & $5 \times 10^{10} ~\GeV$ & $10^{13} ~ \GeV$ & $3 \times 10^{9} ~ \GeV$
        \\
    \hline
    A-2 & $10^3 H_{\mathrm{inf}}$  & $5 \times 10^{10} ~\GeV$ & $10^{7} ~ \GeV$ & $1 \times 10^{7} ~ \GeV $
    \\
      \hline
    B-1 & $\Mpl$  & $5 \times 10^{13} ~ \GeV$ & $10^{16} ~ \GeV$ & $3 \times 10^{11} ~ \GeV $
    \\
      \hline
    B-2 & $\Mpl$  & $5 \times 10^{13} ~ \GeV$ & $10^{10} ~ \GeV$ & $1 \times 10^{9} ~ \GeV $
    \\
      \hline
    \end{tabular}
        \end{center}
    \caption{Four pairs of parameters. A-1 and B-1 are classified as case~1, and A-2 and B-2 are examples of case~2.}
\label{Table. four cases}
\end{table}
\begin{figure*}[htbp]
    \centering
    \includegraphics[width=\hsize]{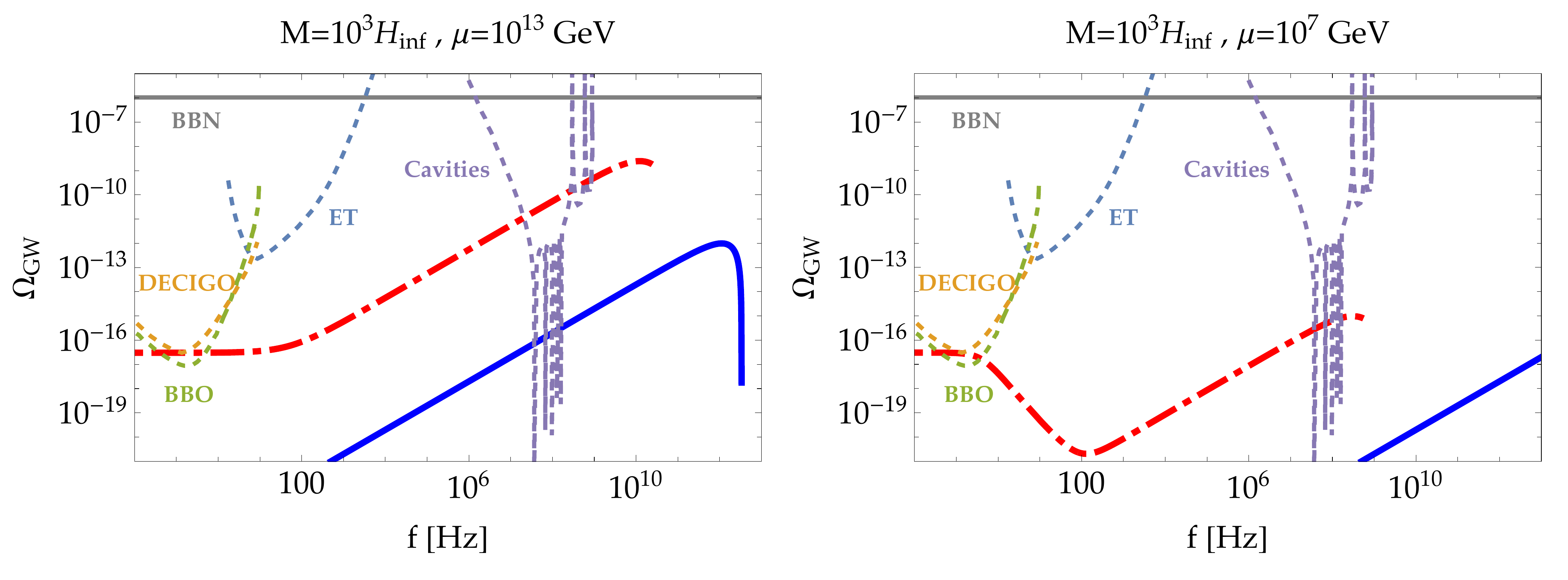}
    \caption{
The current density parameter of GWs with $M = 10^3 H_{\mathrm{inf}}$ and $\mu=10^{13}~\GeV$ (left) or $\mu=10^{7}~\GeV$ (right). The red dash-dotted line denotes the GWs originating from inflation, while the blue solid line is the one from the graviton Bremsstrahlung. The gray line is an upper bound $h^2 \Omega_\GW < 1.3\times 10^{-6}$ to have the successful BBN~\cite{Yeh:2022heq}. Dashed lines smaller than MHz exhibit the future sensitivity prospects by DECIGO, Einstein Telescope, and the Big Bang Observer (taken from Ref.~\cite{DEramo:2019tit}). The GHz region can be probed by gravitational wave detectors based on resonant cavities~\cite{Herman:2020wao, Herman:2022fau}.
    }
    \label{fig: GW Hubble0001}
\end{figure*}
\begin{figure*}[htbp]
    \centering
    \includegraphics[width=\hsize]{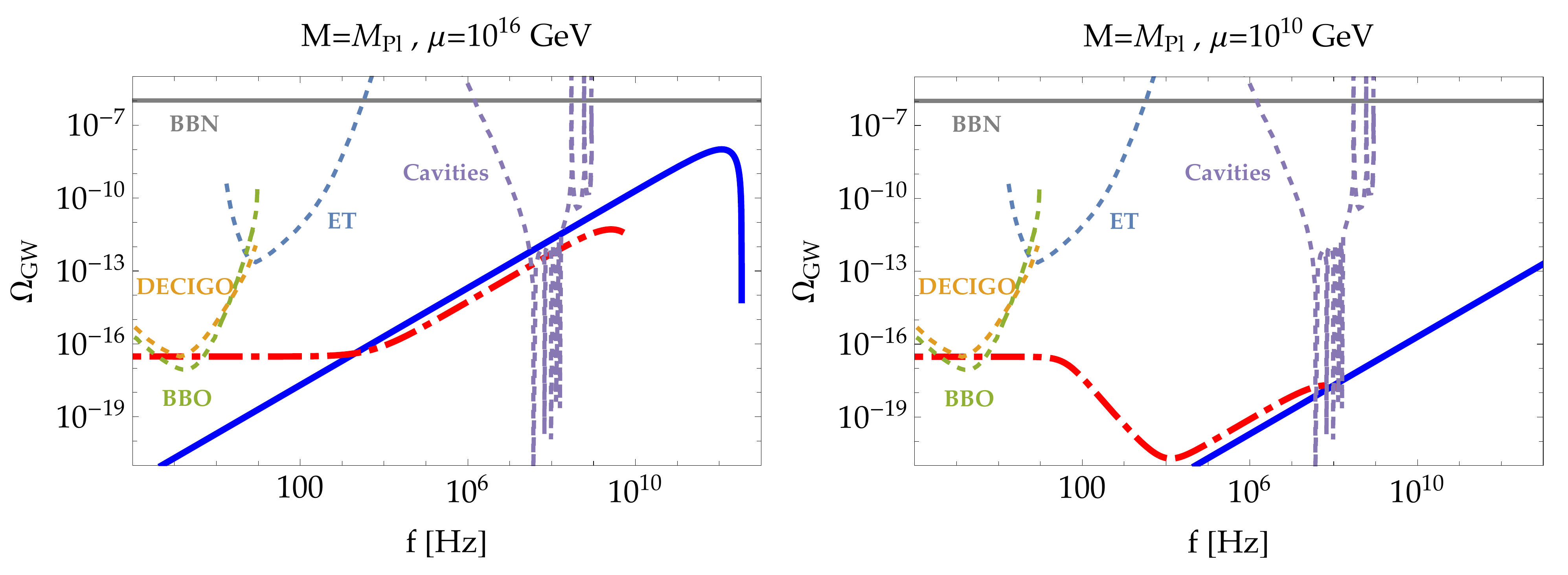}
    \caption{
    Same as Fig.~\ref{fig: GW Hubble0001} but for the case with $M = \Mpl$ and $\mu=10^{16}~\GeV$ (left) or $\mu=10^{10}~\GeV$ (right).
    }
    \label{fig: GW Mpl}
\end{figure*}

 Fig.~\ref{fig: GW Hubble0001} shows the current density parameter of GWs with $M = 10^3 H_{\mathrm{inf}}$. The left panel corresponds to the case~1 where the inflationary GWs behave in an ordinary way and we can see that their amplitude is much greater than GWs from the graviton Bremsstrahlung. 
 In the right panel corresponding to the case~2, we see that the inflationary GWs decrease around $100$ Hz due to the presence of the intermediate matter-dominated universe. Accordingly, in the GHz region, the predicted amplitude is relatively small compared to the case~1. However, GWs from the graviton Bremsstrahlung are also suppressed for small $\mu$, so that they are buried in inflationary GWs.

If the parent particles can be as heavy as the Planck scale, the situation can be different as the suppression of GWs from the graviton Bremsstrahlung becomes small. This can be seen in Fig.~\ref{fig: GW Mpl}. The left panel, with $\mu = 10^{16} ~\GeV$ (case 1), shows that GWs from the graviton Bremsstrahlung surpass the ones from inflation. We check that this relation between the two magnitudes holds when $10^{16}~\GeV \lesssim \mu \lesssim \Mpl$ and they are comparable when $10^{14}~\GeV \lesssim \mu \lesssim 10^{15}~\GeV$. 
The right panel of Fig.~\ref{fig: GW Mpl}, with $\mu = 10^{10} ~\GeV$ (case 2), shows that GWs from inflation would be greater than ones from the graviton Bremsstrahlung. This relation holds true in almost all parameter space of the case~2, say $10^{-2}~\GeV \lesssim \mu \lesssim 10^{13}~\GeV$. It may be interesting to look at modes that did not cross the horizon as the inflationary GWs decay exponentially~\cite{Pi:2024kpw} while the one from the graviton Bremsstrahlung still grows linearly in $f$ in the GHz frequency range. However, we have to note that the frequency region that can be probed by resonant cavities is narrow and a fine-tuning of the parameter is needed.
To add more, GWs from the graviton Bremsstrahlung for $f < f_{\rm peak}$ have the same power-law behavior as the inflationary GWs which reenter the horizon in the kination phase. To distinguish GWs from two sources, multi-frequency observations, e.g., using the deciHz (DECIGO/BBO) and GHz (Cavities) bands, would be important.
\section{Conclusions}\label{sec. discussions}
In this paper, we investigate the detectability of GWs from the inevitable graviton Bremsstrahlung process in the presence of the kination phase that relatively amplifies inflationary GWs at higher frequencies due to the quickly decaying background energy density. Figs~\ref{fig: GW Hubble0001} and~\ref{fig: GW Mpl} show the predictions of the current density parameter of GWs with fixed $M$ and $\mu$ in comparison with the sensitivity curves of some future detectors. We find that GWs from the graviton Bremsstrahlung are buried in ones originating from inflation if the parent particles are not as heavy as the Planck scale. If the mass of the parent particles is close to the Planck scale, the graviton Bremsstrahlung process can produce a sizable amount of GWs, surpassing the primordial ones.

Our analysis adopts the estimations studied in Ref.~\cite{Hashiba:2018iff}, where the parent particles couple conformally to gravity. While we expect that our setup is quite general for inflation models that undergo the kination phase and reheat the universe gravitationally, it would be interesting to study thoroughly how the breaking of conformal symmetry affects the amount of the energy density at the production, the mass of produced particles, and the spectrum of GWs. It would also be intriguing to consider realistic models that fall into the category discussed in our work. 
\section*{Acknowledgments}
R.I. is supported by JST SPRING, Grant Number JPMJSP2125, and the ``Interdisciplinary Frontier Next-Generation Researcher Program of the Tokai Higher Education and Research System''. Y.M. is supported by JSPS KAKENHI Grants No.~JP22KJ1600.
S.Y. is supported by JSPS KAKENHI Grants No. JP20K03968, JP24K00627, and JP23H00108. 

\bibliography{Bib}
\bibliographystyle{utphysYM}
\end{document}